\documentclass[12pt]{article}

\usepackage{cite}
\usepackage{color}
\usepackage{amsmath}
\usepackage{amsfonts}
\usepackage{amssymb}
\usepackage{caption}
\usepackage{graphicx}
\usepackage{slashed}            
\usepackage{subfig}             
\usepackage{xspace}				

\usepackage[english]{babel}
\usepackage{fancyhdr}
\usepackage{psfrag}
\usepackage[applemac]{inputenc}
\usepackage{wrapfig}

\newcommand{\drawsquare}[2]{\hbox{%
\rule{#2pt}{#1pt}\hskip-#2pt
\rule{#1pt}{#2pt}\hskip-#1pt
\rule[#1pt]{#1pt}{#2pt}}\rule[#1pt]{#2pt}{#2pt}\hskip-#2pt
\rule{#2pt}{#1pt}}

\newcommand{\Yfund}{\drawsquare{7}{0.6}}

\newcommand{\arXiv}[2]{\href{http://arxiv.org/pdf/#1}{{\tt #2/#1}}}
\newcommand{\arXivold}[1]{\href{http://arxiv.org/pdf/#1}{{\tt #1}}}

\renewcommand{\tilde}{\widetilde} 

\newcommand{\TeV}{\,\mathrm{TeV}}
\newcommand{\GeV}{\,\mathrm{GeV}}

\newcommand{\beq}{\begin{eqnarray}}
\newcommand{\eeq}{\end{eqnarray}}

\newcommand{\Lslash}[1]{\slash{ \hspace{-2.5mm} #1}}

\newcommand{\eq}[1]{Eq.~(\ref{#1})}
\newcommand{\fig}[1]{Fig.~\ref{#1}}

\newcommand{\bag}{\begin{align}}
\newcommand{\eag}{\end{align}}

\newcommand{\ie}{$\textnormal{i.e.}$ }

\newcommand{\order}[1]{O(#1)}

\newcommand{\nn}{\nonumber}

\newcommand{\Lag}{\mathcal{L}}

\newcommand{\BR}{\mathrm{BR}}

\usepackage[hypertexnames=false]{hyperref}		

\begin{document}
\begin{titlepage}

\vskip.5cm

\begin{center} 
{\huge \bf A Higgslike Dilaton } 
\end{center}

\begin{center} 
{\bf \  Brando Bellazzini$^{a, b}$, Csaba Cs\'aki$^c$, Jay Hubisz$^d$, Javi Serra$^c$ and John Terning$^e$} \\
\end{center}
\vskip 8pt

\begin{center} 
$^{a}$ {\it  Dipartimento di Fisica e Astronomia, Universit\`a di Padova and INFN, Sezione di Padova, Via Marzolo 8, I-35131 Padova, Italy} \\

\vspace*{0.1cm}

$^{b}$ {\it SISSA, Via Bonomea 265, I-34136 Trieste, Italy} \\

\vspace*{0.1cm}

$^{c}$ {\it Department of Physics, LEPP, Cornell University, Ithaca, NY 14853, USA} \\

\vspace*{0.1cm}

$^{d}$ {\it  Department of Physics, Syracuse University, Syracuse, NY  13244} \\

\vspace*{0.1cm}

$^{e}$ {\it Department of Physics, University of California, Davis, CA 95616} \\

\vspace*{0.1cm}

{\tt  
 \href{mailto:brando.bellazzini@pd.infn.it}{brando.bellazzini@pd.infn.it},
\href{mailto:csaki@cornell.edu}{csaki@cornell.edu}, \\
 \href{mailto:jhubisz@phy.syr.edu}{jhubisz@phy.syr.edu},  
 \href{mailto:js993@cornell.edu}{js993@cornell.edu},
 \href{mailto:jterning@gmail.com}{jterning@gmail.com}}

\end{center}

\vglue 0.3truecm

\centerline{\large\bf Abstract}
\begin{quote}

We examine the possibility that the recently discovered 125 GeV higgs-like resonance actually corresponds to a dilaton: the Goldstone boson of scale invariance spontaneously broken at a scale $f$.  Comparing to LHC data  we find that a dilaton can reproduce the observed couplings of the new resonance as long as $f\approx v$, the weak scale. This corresponds to the dynamical assumption that only operators charged under the electroweak gauge group obtain VEVs. The more difficult task is to keep the mass of the dilaton light compared to the dynamical scale, $\Lambda \sim 4 \pi f$, of the theory. In  generic, non-supersymmetric theories one would expect the dilaton mass to be similar to  
$\Lambda$. The mass of the dilaton can only be lowered at the price of some percent level (or worse) tuning and/or additional dynamical assumptions: one needs  to suppress the contribution of the condensate to the vacuum energy (which would lead to a large dilaton quartic coupling), and to allow only almost marginal deformations of the CFT.

\end{quote}

\end{titlepage}

\newpage

\setcounter{equation}{0}
\setcounter{footnote}{0}

\section{Introduction}

The recent discovery of a higgs-like resonance at 125 GeV by the CMS~\cite{CMSHiggs} and ATLAS~\cite{AtlasHiggs} experiments at the LHC strongly disfavors technicolor/higgsless models, where electroweak symmetry (EWS) is broken via strong dynamics, without the presence of a light, narrow higgs-like state. The only escape for such models would be if the strong dynamics (simultaneously with breaking EWS) also produced a higgs-like resonance which is light and narrow for some reason unrelated to EWS breaking. While this possibility may at first sound contrived, there is a well-motivated case that could actually fulfill these requirements. The crucial realization is that the properties of the Standard Model (SM) higgs boson are largely fixed by the approximate conformal invariance in the limit when the higgs potential is turned off. In this case the higgs VEV is arbitrary (it is a flat direction), and its value at $v=246$ GeV will spontaneously break the approximate conformal symmetry as well as EWS. In this scenario the higgs is identified with the massless dilaton \cite{Ellis:1975ap} of conformal breaking, with the conformal symmetry breaking scale $f=v$. Thus the higgs couplings are mainly dictated by this conformal invariance. If the technicolor/higgsless theory itself was approximately conformal, then it would be plausible that the condensate breaking EWS also spontaneously broke conformal invariance, and the resulting dilaton could have properties very similar to that of the SM higgs \cite{Yamawaki,Bardeen,Clark,Dzhikiya,Appelquist,Grinstein,Yamawakinew,CHL,Witek,Vecchi,Barger,Ryskin,Foot,Campbell,Coleppa,Elander:2012fk}. 

The aim of this paper is two-fold: to examine  the phenomenology of such a dilaton, and
to investigate the dynamics that could be responsible for producing a light dilaton. First we will show that at this point a light dilaton could still easily be in agreement with the very preliminary observed values of the branchings fractions of the higgs-like resonance. For this we derive the dilaton couplings to matter using symmetry arguments. The dilaton couplings are generically $v/f$ suppressed compared to SM higgs couplings. Fermion couplings are also modified by the anomalous dimensions of the SM fermions. Couplings to massless gauge bosons are loop induced and are determined by the $\beta$ function coefficients of the composites. Our approach differs somewhat from the usual assumptions made about embedding the SM into a conformal theory: it is commonly assumed that the entire SM appears as composites of the broken conformal sector~\cite{Witek}, which would imply a dilaton coupling to gluons that is much too large and already disfavored by the LHC data~\cite{Low:2012rj,Strumia,Elander:2012fk,Montull:2012ik} \footnote{Reference \cite{Ellis:2012hz} fits the dilaton couplings assuming only an overall rescaling of the SM higgs couplings.}. Instead, we will follow the route of partial compositeness \cite{Kaplan:1991dc} as in realistic warped extra dimensional models~\cite{CHL,ADMS}, where both the SM fermions and gauge bosons are mixtures of elementary and composite states. 
We emphasize that the coupling to the $W$ and $Z$ can only be made realistic and consistent with electroweak precision tests if the scale of conformal breaking is within about $10 \%$ of the scale of EWSB: $f\approx v$. This is achieved in models where only operators charged under the electroweak gauge group obtain a VEV. 
 
In the second part of the paper we investigate the question of how plausible it would be to obtain a light dilaton in a strongly coupled theory without fine tuning. We emphasize that the main difference between Goldstone bosons of ordinary compact internal symmetries and the dilaton is that a non-derivative quartic scalar self-interaction term for the dilaton is allowed by the conformal symmetry. Applying naive dimensional analysis (NDA) for the dilaton effective action one finds that the natural value of this quartic is large, of order $16 \pi^2$, which will make it very difficult to produce stable, spontaneous conformal symmetry breaking: to stabilize this potential at finite (but hierarchically small values) of the scales $f$ one generically needs a very large explicit breaking term, which in turn will produce a large non-suppressed mass for the dilaton. This is what happens in QCD-like models \cite{Holdom,Kutasov} or in simple walking technicolor theories: the coupling responsible for producing the condensate is the strong coupling itself, which will need to flow to large values and thus have a large $\beta$ function, which implies a large explicit breaking and thus a heavy dilaton. In theories of this sort $m_{dil}\sim \Lambda$, where $\Lambda$ is the scale of strong coupling, as expected. A conclusion different from this can be obtained only if the quartic, allowed by the symmetries, is significantly smaller than its natural NDA value, effectively starting out with a flat direction, which corresponds to a tuning in the theory. In this case an explicit breaking by an almost-marginal operator of dimension $4-\epsilon$  with a perturbative coupling might stabilize the dilaton at a hierarchically small VEV.\footnote{This is the scenario explored in ref. \cite{postmodern}.} Then the mass of the dilaton can also be suppressed by the anomalous dimension of the perturbing operator $\epsilon$: the smaller the explicit breaking the smaller the dilaton mass is expected to be. This is nothing but the generic story behind the Goldberger-Wise (GW) radion stabilization mechanism \cite{Goldberger:1999uk} in the Randall-Sundrum (RS) warped extra dimensional model \cite{Randall:1999ee} as explained in~\cite{RattazziZaffaroni}. Of course supersymmetry can be used to produce flat directions and thus to forbid the leading quartic scalar self-coupling, in which case a naturally light dilaton can be readily obtained, however supersymmetry plays an essential role in those models. 

The paper is organized as follows. In Section 2 we review the basic properties of a dilaton, and explain the assumptions of the setup including a composite and an elementary sector. We then show how to use a spurion analysis to obtain the dilaton couplings to the various composite and elementary fields. In Section 3 we specify the couplings in the case of a partially composite SM sector, paying special attention to the couplings to massless gauge fields. We present the dilaton effective Lagrangian relevant for LHC studies in Section 4, together with constraints both from LHC higgs results and electroweak precision measurements. In Section 5 we present the general discussion of the conditions necessary for obtaining a light dilaton, and estimate the amount of tuning required. Section 6 contains a supersymmetric toy model that has a naturally light dilaton, while in Section 7 we study the Goldberger-Wise stabilized Randall-Sundrum model. Finally we conclude in Section 8.

\section{Scaling and Dilaton basics}
\label{sec:basics}
\setcounter{equation}{0}

In this section we summarize the basic properties of scale transformations and  dilaton couplings.   Scale transformations~\cite{Coleman} are given by (for $x\to x' = e^{-\alpha} x$)
\begin{equation}
 {\cal O} (x)\to {\cal O}'(x)=e^{\alpha \Delta} {\cal O} (e^\alpha x) \, ,
\end{equation}
where $\Delta$ is the matrix of dimensions (including classical and quantum effects) for the operators ${\cal O}$. The action changes under scale transformations as
\begin{equation}
S=\sum_i \int d^4 x\, g_i \mathcal{O}_i(x)\longrightarrow S^\prime= \sum_i \int d^4 x e^{\alpha(\Delta_i-4)}g_i\mathcal{O}_i(x) \, ,
\end{equation}
which implies the well-known result that all operators must have dimension $\Delta_i =4$ for all $\mathcal{O}_i$ in order for the action to be scale invariant. 
The linearized transformation of the action is then 
\begin{equation}
S\longrightarrow S+\sum_{i}\int d^4x\, \alpha g_{i}(\Delta_{i}-4)\mathcal{O}_i(x) \,.
\label{eq:scaletrans}
\end{equation}

Let us assume  that scale invariance is broken spontaneously by the VEV of a  dimensionful operator $\langle \mathcal{O}\rangle=f^n$ where $n$ is the classical dimension of $\mathcal{O}$. The spontaneous breaking of scale invariance (SBSI) will imply the existence of a  Goldstone boson for scale transformations, the dilaton, which transforms inhomogeneously under scale transformations:
\begin{equation}
\sigma(x)\to \sigma(e^\alpha x)+ \alpha f \, .
\end{equation}
The low-energy effective theory can be obtained by replacing the VEV with the non-linear realization
\begin{equation}
f\to f\, \chi \equiv f \, e^{\sigma /f} \, ,
\label{eq:replacement}
\end{equation}
and requiring that it is invariant under scale transformations:
\beq
{\cal L}_{eff}&=&  \sum_{n, m\geqslant 0} \frac{a_{n,m}}{(4 \pi)^{2(n-1)}\,f^{2(n-2)}} \frac{\partial^{2n}\chi^m  }{\chi^{2n+m-4}} \\
&=&- a_{0,0} \, (4\pi)^2 f^4 \chi^4 +\frac{f^2}{2}(\partial_\mu \chi)^2  +  \frac{a_{2,4}}{(4 \pi)^{2} }  \frac{(\partial \chi )^{4} } {\chi^4} + \ldots
\eeq
where $a_{n,m} \sim O(1)$, and $a_{1,1}=1/2$ corresponds to canonical normalization, and $a_{2,4}$ is determined by the proof of the $a$-theorem \cite{Komargodski}.
The complete set of dilaton couplings within the scale-invariant sector can be obtained by the replacement in (\ref{eq:replacement}). However, a more systematic way is to 
take advantage of the (approximate) scale invariance of the Lagrangian at high energies,
in order to build an effective Lagrangian for energies below $\Lambda \sim 4 \pi f$ where scale invariance is preserved by means of insertions of the dilaton field as defined in \eq{eq:replacement}.

The general assumption we will be making is that there is a conformal sector which is spontaneously broken, which we will refer to as the ``composite sector", and that there is another sector weakly coupled to it that explicitly breaks the conformal invariance, which we will refer to as the ``elementary sector". There could also be small explicit breaking terms within the composite sector. The SM matter fields will be mostly elementary, but some of them (for example the top) can be partly composite.

\subsection{Composite sector couplings}
\label{sec:compocoupling}

Let us assume that in the UV the theory is determined by the Lagrangian 
\begin{equation}
{\cal L}_{CFT}^{UV} =\sum_i g_i {\cal O}^{UV}_i \, ,
\end{equation}
where the operators above include both scale invariant ($\Delta_i^{UV}=4$) and small explicit breaking terms ($\Delta_i^{UV}\neq 4$). We treat the explicit breaking couplings as spurions under  scale transformations, and assign to them a fictitious scaling dimension
\begin{equation}
\left[ g_i\right] =4-\Delta_i^{UV} \, .
\end{equation}
The low-energy effective theory, valid below the scale $\Lambda \sim 4\pi f$, might present a different field content.
The Lagrangian can be written as
\beq
{\cal L}_{CFT}^{IR}=\sum_j c_j \left(\Pi g_i^{n_i} \right) {\cal O}^{IR}_j  \chi^{m_j} \,,
\eeq
where $c_j$ is an unknown function of the scale invariant couplings and we have expanded in the small explicit breakings. The power of $\chi$ is determined by requiring scale invariance:
\beq
m_j={4- \Delta^{IR}_j - \sum_i n_i(4-\Delta_i^{UV})} \,.
\eeq
For terms with a single power of a symmetry breaking coupling and to leading order in the dilaton field we have
\begin{equation}
{\cal L}^{IR}_{breaking}=\sum_{j} \, c_j \,g_i \,\left(\Delta_i^{UV}-\Delta^{IR}_j \right) \mathcal{O}_j^{IR}  \,\frac{\sigma}{f} \,.
\label{eq:compositecouplings}
\end{equation}
For terms involving no explicit symmetry breaking we have
\begin{equation}
{\cal L}^{IR}_{symmetric}=\sum_{j} \,  c_j \, \left(4-\Delta^{IR}_j \right) \mathcal{O}_j^{IR} \, \frac{\sigma}{f}    \,.
\label{eq:compositecouplings2}
\end{equation}
This is just the well-known special case that the dilaton couples to the trace of the energy-momentum tensor
\begin{equation}
{\cal L}_{eff}=-\frac{\sigma}{f}\, \mathcal{T}^\mu_\mu \, .
\end{equation}
For instance, the trace anomaly is included in the perturbative contribution to $\Delta^{IR}_j$
for $\mathcal{O}_j^{IR} = -(F_{\mu\nu})^2/(4g^2)$
\beq
4-\Delta^{IR}_j= 2 \gamma(g)=  \frac{2\beta(g)}{g} \,.
\eeq
This is, for example, the case in the original RS  model, where the entire SM is assumed to live in the IR brane, thus being fully composite. 

\subsection{Elementary-Composite couplings}
\label{sec:mixings}

The more interesting cases however do not correspond to a SM fully embedded into the composite sector. Instead, the SM fields are considered external to the CFT dynamics, 
as in the realistic RS-model of Agashe et al \cite{ADMS}. Indeed the couplings of the dilaton in this case do not quite follow the above results as shown in~\cite{CHL}, since the elementary fields introduce explicit breakings of the conformal invariance through their (weak) couplings to the composite operators. In this case one has to perform a spurion analysis to derive the couplings of the dilaton. 
Let us consider that at high energies the Lagrangian can be written as
\begin{equation}
\label{Selem}
{\cal L}^{UV}={\cal L}_{CFT}^{UV}+{\cal L}_{elem}+\sum_i y_i \,O_{elem, i}\, {\cal O}^{UV}_{CFT, i} \, ,
\end{equation}
where the elementary-composite interactions generically break conformal invariance since their dimensions are not four. Following the spurion analysis
\begin{equation}
\left[ y_i \right] =4-\Delta_{elem,i}^{UV} -\Delta^{UV}_{CFT,i} \, .
\end{equation}
Notice that the scaling dimensions of the elementary fields might differ from their classical dimensions. This is due to the CFT contribution to their wave function renormalization. This is generically a subleading effect, unless the coupling $y_i$ gets strong, or in special cases like gauge fields, where gauge invariance fixes the couplings.

The low-energy effective Lagrangian then takes the form
\beq
{\cal L}_{eff}={\cal L}^{IR}_{CFT}+{\cal L}_{elem}+\sum_j c_j\, y_i \,O_{elem,i} \, {\cal O}^{IR}_{CFT,j} \, \chi^{m_j} +{\cal O}(y^2) \, ,
\eeq
The power of $\chi$ is determined by requiring scale invariance:
\beq
m_j&=&4- \Delta^{IR}_{CFT,j} -\Delta_{elem,i}^{IR}- \left(4-\Delta_{elem,i}^{UV} -\Delta^{UV}_{CFT,i}\right)
\\
&=&\Delta_{elem,i}^{UV} -\Delta_{elem,i}^{IR}+\Delta^{UV}_{CFT,i} - \Delta^{IR}_{CFT,j}  ~.
\eeq
This analysis can be easily extended to include terms of higher order in $y_i$.
Notice that in general the dilaton also couples to operators build only with elementary fields, if the proper powers of $y_i$ are introduced.
Likewise, dilaton couplings to composite operators will be generated, as explained in Section~\ref{sec:compocoupling}.

\subsubsection*{Partially composite fermions}

Consider the following interaction of the elementary fermions $\psi_L$, $\psi_R$ with composite operators $\Theta_L$, $\Theta_R$ at high energies
\beq
{\cal L}_{int}= y_L \psi_L \Theta_R+ y_R \psi_R \Theta_L + h.c. \,.
\label{eq:psimixing}
\eeq
These realize the paradigm of partial compositeness \cite{Kaplan:1991dc,ADMS}, in which the flavor structure of the SM is reproduced at low energies by fixing the amount of mixing $y_L$, $y_R$ and the dimensions of $\Theta_L$, $\Theta_R$ for each SM chiral fermion.
The spurious scaling dimensions are
\beq
\left[ y_L\right]= 4- \Delta^{UV}_{\psi_L}- \Delta^{UV}_{\Theta_R} \,, \quad
\left[ y_R\right]= 4- \Delta^{UV}_{\psi_R}- \Delta^{UV}_{\Theta_L} \,.
\eeq
After integrating out the massive composite degrees of freedom, the following interaction is generated
\beq
{\cal L}_{eff}= -M \,y_L\, y_R \,\psi_L \psi_R \chi^m +h.c. \, ,
\eeq
where 
\beq
m&=& 4-\left(4- \Delta^{UV}_{\psi_L}- \Delta^{UV}_{\Theta_R}+4- \Delta^{UV}_{\psi_R}- \Delta^{UV}_{\Theta_L}\right)- \Delta^{IR}_{\psi_L}- \Delta^{IR}_{\psi_R} \\
&=& \Delta^{UV}_{\psi_L}- \Delta^{IR}_{\psi_L}+ \Delta^{UV}_{\psi_R}- \Delta^{IR}_{\psi_R}+\Delta^{UV}_{\Theta_L}+\Delta^{UV}_{\Theta_R}-4 \, .
\eeq
Using the conventions of AdS/CFT and RS \cite{Cacciapaglia:2008bi}
\beq
\Delta^{UV}_{\Theta_L}=2+c_L \, ,\quad \Delta^{UV}_{\Theta_R}=2- c_R \,,
\eeq
where $c_L > -1/2$ and $c_R < 1/2$.
Neglecting the perturbative anomalous dimensions of the elementary fermions we have the dilaton coupling
\beq
{\cal L}_{eff}&=& -M \, y_L\,  y_R \, \psi_L \psi_R \chi^{c_L-c_R}~.
\eeq
The same result can be obtained by following the dependence on the breaking scale $f$ of the low-energy coupling $y(\mu)$. This follows the renormalization group equation \cite{Agashe:2004rs}
\beq
\frac{d y_{L,R}}{d \ln \mu} = \gamma_{L,R} \, y_{L,R} + O(y_{L,R}^3) \, , \quad \gamma_{L,R} = \pm c_{L,R} -1/2 \, ,
\label{yrun}
\eeq
which determines the low-energy value of $y_{L,R}$,
\beq
y_{L,R}(\mu) \simeq y_{L,R}(\mu_0) \left( \frac{\mu}{\mu_0} \right)^{\gamma_{L,R}} \, .
\eeq
In the low-energy theory the mass term $\psi_L \psi_R$ has a coefficient
$M y_L(f) y_R(f)$ with $M \propto f $ and replacing $f$ by $f e^{\sigma/f}$ we find a linear dilaton coupling
\beq
- m_{\psi} (1+\gamma_L+\gamma_R) \psi_L \psi_R \frac{\sigma}{f} =- m_{\psi} (c_L-c_R) \psi_L \psi_R \frac{\sigma}{f}\, .
\label{eq:sigmapsi}
\eeq
where we have identified $m_\psi = M y_L y_R$.

\subsubsection*{Partially composite gauge fields}
Another simple example: consider the coupling of an elementary gauge field to a global current of the CFT, given by
\beq
{\cal L}= -\frac{1}{4 g_{UV}^2}F_{\mu\nu}F^{\mu\nu}+ A_\mu \mathcal{J}^\mu~.
\label{Amix}
\eeq
Since $\mathcal{J}^\mu$ is a conserved current we assign the spurion dimension
\beq
[g_{UV}]=\Delta_A^{UV}-1\,.
\eeq
The term in the low-energy theory is then
\beq
{\cal L}_{eff}= -\frac{1}{4 g^2} F_{\mu \nu}F^{\mu\nu}\chi^m \, ,
\eeq
where $[g]=[g_{UV}]$ and
\beq
m = 4-2[1+\Delta_A^{IR}]+2[g]= 2(\frac{\beta_{IR}}{g}-\frac{\beta_{UV}}{g}) \, .
\eeq

The same result at one-loop order can be obtained by following the dilaton dependence of the breaking scale $f$.
We can write the IR gauge coupling as
\beq
\frac{1}{g^2(\mu)}= \frac{1}{g^2(\mu_0)} - \frac{b_{UV}}{8 \pi^2} \ln \frac{\mu_0}{f} - \frac{b_{IR}}{8 \pi^2} \ln \frac{f }{\mu} \, ,
\eeq
In the low-energy theory $F^{\mu \nu}F_{\mu\nu}$ has a coefficient
$ -1/4g^2(\mu)$ and replacing $f$ by $f e^{\sigma/f}$ we find a linear dilaton coupling
\beq
\frac{g^2}{32 \pi^2} \left( b_{IR}-b_{UV}\right) F^{\mu \nu}F_{\mu\nu} \frac{\sigma}{f}~.
\label{eq:sigmaF}
\eeq
In the next section we will specify these formulae for the interesting case when the SM is partially embedded into the conformal sector, but most of the SM fields remain elementary.

\section{Dilaton phenomenology}
\label{sec:pheno}
\setcounter{equation}{0}

In this section we discuss the couplings of the dilaton to SM fields, in order to eventually address its viability as a candidate for the recently discovered higgs-like resonance at 125 GeV.

We assume that SM fermions (except perhaps the top) and gauge bosons are elementary with weak couplings to the strong sector. This assumption is supported by mounting experimental evidence on the elementary nature of the SM leptons and light quarks, as well as of the photon, gluon, and transverse components of the $W$ and $Z$.
On the other hand, the dilaton and the longitudinal components of the $W$ and $Z$ are assumed to be composites. This is equivalent to the assumption that EWSB is fully driven by the composite sector. Furthermore, the heaviness of the top quark might be an indication of its compositeness as well, a possibility yet to be tested at the LHC. Here we will consider a fully composite right-handed top.

Next we use the general results of the previous section to specify couplings of the elementary SM fields to the strong sector. The coupling of a gauge field $A_\mu$ is dictated by gauge invariance: it couples linearly to  the corresponding global current $\mathcal{J}^\mu$ of the strong sector with the gauge coupling $g$.  For a chiral fermion, $\psi$, we will assume that it couples linearly to a (single) fermionic composite operator $\Theta$ with strength $y$. These interactions realize the framework of partial compositeness for the SM. The corresponding low-energy couplings of the dilaton to the SM fields can be derived following the discussion in Section~\ref{sec:mixings}, and they are given by \eq{eq:sigmapsi} and \eq{eq:sigmaF} respectively for fermions and gauge fields.
For completeness, let us write down the full effective Lagrangian for the SM at low energies and derive the couplings of the dilaton at leading order in $1/f$. For operators with classical dimension less than or equal to four, the Lagrangian can be divided into kinetic terms,
\beq
\Lag_{kin} = - \frac{1}{4 g_A^2} \left( F_{\mu \nu}^{(A)} \right)^2 + i \bar{\psi} \Lslash{D} \psi \, ,
\label{eq:Lkin}
\eeq
and mass terms,%
\beq
\Lag_{mass} = \frac{v^2}{2} |D_\mu \Sigma|^2 -  Y_{\psi} \frac{v}{\sqrt{2}} \psi_L \Sigma \psi_R + h.c. \, 
\label{eq:Lmass}
\eeq
Mass terms arise after the spontaneous breaking of scale and electroweak invariances.
The first term, giving rise to the $W$ and $Z$ masses, is just the leading operator in derivatives of the non-linear sigma model parametrized by $\Sigma$, which contains the Goldstone bosons of the spontaneous breaking of EWS, $SU(2)_L \times SU(2)_R \to SU(2)_V$.
The second term comes from a $SU(2)_V$ invariant strong sector operator of the form $\bar{\Theta}_{L} \Theta_{R}$, after rotation of the fermionic mixing term in \eq{eq:psimixing}.
Therefore we expect that the Yukawa couplings scale as $Y_\psi \propto y_L y_R$ at leading order in $y_{L,R}$.

The couplings of the dilaton to the SM fields are then given by
\beq
\delta \Lag_{kin} = \frac{g_A^2}{32 \pi^2} \left( b_{IR}^{(A)} - b_{UV}^{(A)} \right)
\left( F_{\mu \nu}^{(A)} \right)^2 \frac{\sigma}{f} \, ,
\label{deformkin}
\eeq
and
\beq
\delta \Lag_{mass} = \left( 2 m_W^2 W^+_\mu W^{-\mu} + m_Z^2 Z_\mu^2 \right) \frac{\sigma}{f} -
Y_\psi \frac{v}{\sqrt{2}} \psi_L \psi_R (1+\gamma_{L} + \gamma_{R}) \frac{\sigma}{f} + h.c. \, ,
\label{deformmass}
\eeq
where we have moved to unitary gauge and canonically normalized the gauge fields.
The $\beta$ function coefficient $b_{UV}^{(A)}$ and $b_{IR}^{(A)}$ parametrize the explicit breaking of scale invariance in the UV and the IR due to the contribution of composite fields to the running of the gauge coupling $g_A$. The anomalous dimensions $\gamma$ measure the explicit breaking associated to the mixing in the UV between elementary and composite fields. 
Following the arguments of Section~\ref{sec:mixings} on the generation of fermion masses, we expect an enhancement of the couplings to light fermions, since these require $\gamma > 0$. 
However, the generation of flavor structure is a model dependent subject, and this expectation might not be satisfied. In fact, as we discuss below, it might be required for the strong sector to be flavor symmetric and therefore to have equal $\gamma$'s among generations.
Furthermore, it might be possible that the strong sector yields negative anomalous dimensions at low energies, $\gamma < 0$, and thus reduced dilaton couplings to fermions even if $v \approx f$.

Notice that if $\gamma$'s are flavor dependent, the mass terms and the dilaton Yukawa couplings cannot be diagonalized simultaneously\footnote{For simplicity we assumed that $\Delta^{UV}_{\psi}-\Delta^{IR}_{\psi}\simeq 0$}
\begin{equation}
y^i_{L\,a} y^j_{R\,b}\Sigma^{ab}\frac{v}{\sqrt{2}}\psi_L^{i}\psi_{R}^j\left[1+\frac{\sigma}{f}\left(1+\gamma^a_L+\gamma^b_L \right)+\ldots\right]
\end{equation}
inducing dilaton mediated flavor changing neutral currents. The matrix $\Sigma^{ab}=\langle\Theta^a_L\Theta_R^b\rangle$ arises from integrating out strong sector fields $\Theta^a_{L,R}$.
After passing to the mass basis one could conveniently parametrize flavor violating interaction as in \cite{Azatov:2008vm}
\begin{equation}
\psi_L^{i}\psi_{R}^j\left[m_i \left(1+\frac{\sigma}{f} \right) \delta_{ij}+a_{ij}\sqrt{m_i m_j}\frac{\sigma}{f}+\ldots\right]\,.
\end{equation}
and generate dangerous tree-level 4-fermion operators of size $\sim a_{ij}\sqrt{m_i m_j}/(m_{dil}^2 f^2)$.
In order to pass the strong bounds on such operators in the quark sector \cite{Bona:2007vi} we demand that the composite sector has an $SU(3)_q \times SU(3)_d \times SU(2)_u$ flavor symmetry. We do not introduce a flavor symmetry for
the right-handed top quark, which is a safe flavor assumption \cite{Redi:2012uj}. In addition, the $t_R$ might be a fully composite field, in which case there is no explicit breaking associated to it, and we can effectively set $\gamma_{t_R} = 0$.
It is remarkable that in the lepton sector the low-energy constraints allow sizable branching ratios (BRs) (up to $\mathcal{O}(10\%)$) into flavor violating decays such as $\sigma\rightarrow \tau\mu, \tau e$ \cite{Blankenburg:2012ex,Harnik:2012pb}.

By taking the limits $f \to v$ and $\gamma_i \equiv \gamma_L + \gamma_R, \, b_{UV}^{(A)} - b_{IR}^{(A)} \to 0$, 
the couplings of the SM higgs to gauge bosons and fermions are reproduced by the dilaton.
We comment on the feasibility of these limits below.

Notice that the scale $f$ of the SBSI must ``contain'' the electroweak scale $v$, that is $f \geqslant v$,
since one of our initial assumptions is that the composite sector is solely responsible for the breaking of EWS.
We can define a \emph{minimal} strong sector as one where $f = v$, that is, all the SBSI carries the electroweak quantum numbers of the higgs VEV.

\subsection{Dilaton couplings to massless SM gauge bosons}
\label{sec:traceanomaly}

From the point of view of LHC phenomenology the most important interactions of the dilaton  are those to massless SM gauge bosons.
Standard lore has that they differ enormously from the SM higgs couplings. 
For example, let us consider the coupling to gluons. 
In the SM it is due entirely to the contribution from the top quark. 
This is easy to understand from the perspective of the higgs low-energy theorems. 
The running of QCD gauge coupling $g_s$ tells us that the kinetic term for the gluons is given by
\begin{equation}
-\frac{1}{4 g_s^2 (\mu) } G_{\mu\nu}^2 \, ,
\end{equation}
with
\begin{equation}
\frac{1}{g_s^2 (\mu)} = \frac{1}{g_s^2 (\Lambda)} - \frac{b^{(3)}_{heavy}}{8\pi^2} \log \frac{\Lambda}{m_{heavy}} -  \frac{b^{(3)}_{light}}{8\pi^2} \log \frac{\Lambda}{\mu} \, ,
\label{qcdrun}
\end{equation}
where $g_s(\mu)$ is evaluated at the scale $\mu \sim m_h$.
The different contributions to the running from colored states has been split into the heavy ones, $m_{heavy} > m_h/2$, and the light ones, with their corresponding $\beta$ function coefficients, $ b^{(3)}_{heavy}$ and  $b^{(3)}_{light}$.%
\footnote{Our conventions are such that $\beta = - b g^3 / 8 \pi^2$.}
Presuming that the higgs VEV is solely responsible for $m_{heavy} = Y v$, we obtain the higgs coupling to gluons by substituting $m_{heavy} \to m_{heavy} ( 1 + h/v )$.
From this, we quickly see that the higgs couples to gluons proportionally to the top quark contribution to the QCD $\beta$ function.

In contrast to the SM higgs, the dilaton couples to gluons even before running any SM particles in the loop, through the trace anomaly.
As explained after \eq{deformkin}, the coupling to gluons is proportional to $b_{IR}^{(3)} - b_{UV}^{(3)}$, to be interpreted as the full contribution to the QCD $\beta$ function from loops of composite states only,
where we have split the running into a UV contribution from the CFT above the scale of SBSI, 
$\Lambda>\mu \gtrsim f$, 
and an IR contribution from light composite states below the scale of SBSI, $b^{(3)}_{UV}$ and $b^{(3)}_{IR}$ respectively.
Such a separation between different contributions is convenient since some of the SM fermions could arise as light composite states from the strong sector, for instance the top quark.
The full contribution to the gauge coupling running can then be written as,
\beq
\frac{1}{g_s^2(\mu)} = \frac{1}{g_s^2(\Lambda)} 
- \frac{b^{(3)}_{UV}}{8\pi^2} \log \left( \frac{\Lambda}{f} \right)
- \frac{b^{(3)}_{IR}}{8 \pi^2} \log \left( \frac{f}{\mu} \right)
- \frac{b_{elem}}{8 \pi^2} \log \left( \frac{\Lambda}{\mu} \right)
+ \frac{1}{\tilde{g}_s^2} \, ,
\label{gsrun}
\eeq
where we have included the contribution to the running from the elementary fields, proportional to  $b_{elem}$, and a threshold corrections at the scale $f$, in the form of a tree-level contribution to the gauge coupling, $1/\tilde{g}_s^2$,
which can be interpreted as arising from the mixing of the elementary gluon field and the associated composite.
Most previous work assumed that the whole SM particle content, in particular QCD, is fully embedded into the strong sector \cite{Witek}.
This means, first, that there is no elementary gluon component, $g_s(\Lambda) \to \infty$,
and second, that the $\beta$ function for QCD vanishes above the scale of SBSI, that is $b^{(3)}_{UV} = 0$.
Therefore, in such a naive dilaton scenario one has $b^{(3)}_{IR} = b^{(3)}_{SM}$.
Recalling that the coupling of the dilaton to gluons can be obtained, analogously to the SM higgs, by making the replacement $f \to f\,\chi$ in the expression for $g_s(\mu \sim m_\chi)$, one obtains a much larger coupling than the SM higgs (assuming $f \sim v$), disfavored by data.
Besides, recent LHC dijet data severely constrains the scale of compositeness of the gluon to $\Lambda \gtrsim 5.5 \TeV > 4\pi v$ \cite{Domenech:2012ai}.
In the more realistic case we are considering, the final answer for the dilaton coupling to gluons depends on $b^{(3)}_{UV}$, effectively a free parameter,%
\footnote{It is only constrained by unitarity bounds on the central charge of CFT's \cite{Rattazzi:2010gj}.}
and $b^{(3)}_{IR}$, which under the assumption of a composite $t_R$, is given by $b_{t_R}^{(3)} = - 1/3$,
although this is model dependent.
However, this is not the final result for the coupling of the dilaton to gluons. 
Just as in the case of the SM higgs, particles heavier than $m_\chi/2$, that is the top quark, also contribute at loop level. 
Using the low-energy theorems as an approximation to the computation of the triangle diagrams, we can include this contribution by cutting the corresponding logarithms in the r.h.s.~of  \eq{gsrun} at $\mu = m_t = y_t v$, which gives an extra dilaton coupling after the substitution $v \to v\,\chi$.
Notice then that the full contribution from $t_R$ is obtained from,
\beq
\frac{1}{g_s^2(\mu)} = -\frac{b^{(3)}_{t_R}}{8 \pi^2} \log \left( \frac{f}{y_t v} \right) + \dots
\eeq
which gives no coupling to the dilaton.
If composite, the heavy right-handed top decouples,
which is expected since the contribution of a almost massless composite that turns out to be heavier than the dilaton should not have been included in $b_{IR}^{(3)}$ in the first place.
The final (approximate) result for the dilaton coupling to gluons is then given by
\beq
-(b^{(3)}_{UV} + b^{(3)}_{t_L}) \frac{\alpha_s}{8 \pi} G^2_{\mu \nu} \frac{ \sigma}{f} \, .
\eeq

A similar analysis can be carried out for the coupling of the dilaton to photons.
The result is obtained by making the obvious substitutions, $g_s \to e$, $b^{(3)}_{UV} \to b^{(EM)}_{UV} = b^{(1)}_{UV} + b^{(2)}_{UV}$, and $b^{(3)}_{IR} \to b^{(EM)}_{IR}$,
and including the low-energy standard loop contributions from the top and the $W$.
Now $b^{(EM)}_{IR}$ includes, besides $t_R$, with $N_c \, b^{(EM)}_{t_R} = -8/9$, the NGB's acting as the longitudinal components of $W^\pm$, with $b^{(EM)}_{W^\pm_L} = -1/3$.
These contributions, because $m_W, m_t \gtrsim m_\chi/2$, effectively decouple.
Altogether, the final (approximate) result for the dilaton coupling to photons is given by
\beq
-(b^{(EM)}_{UV} + b^{(EM)}_{W_T^\pm}+ N_c \, b^{(EM)}_{t_L}) \frac{\alpha}{8 \pi} A^2_{\mu \nu} \frac{ \sigma}{f} \, .
\eeq
We properly account for subleading corrections of order $m_\chi^2/ (2 m_t)^2$ and $m_\chi^2/ (2 m_W)^2$ in Section~\ref{sec:coll}.

\section{The dilaton at colliders}
\label{sec:coll}
\setcounter{equation}{0}

In order to make contact with the recent literature on the properties of the higgs-like particle discovered at the LHC, which we denote by $h$, we can make use of the following effective Lagrangian,
\beq
\mathcal{L}_{eff} &=&
c_V \left( \frac{2 m_W^2}{v} W^{+}_\mu W^{-\mu} + \frac{m_Z^2}{v} Z_\mu^2 \right) h \nn \\
&& - c_t \frac{m_{t}}{v} \bar{t} t \, h - c_b \frac{m_{b}}{v} \bar{b} b \, h - c_\tau \frac{m_{\tau}}{v} \bar{\tau} \tau \, h \nn \\
&& + c_g \frac{\alpha_s}{8 \pi v} G^2_{\mu \nu} h + c_\gamma \frac{\alpha}{8 \pi v} A^2_{\mu \nu}
\, ,
\label{Leff}
\eeq
where we are assuming custodial symmetry in the interaction of $h$ with the electroweak gauge bosons, and we have included only couplings to fermions that are relevant for  LHC phenomenology.
\footnote{We are not including a coupling to $Z\gamma$ because of the absence of current measurements.}
The tree-level values of these coefficients in the SM are given by
\beq
c_{V, SM} = c_{t, SM} = c_{b, SM} = c_{\tau, SM} = 1\, , \quad
c_{g,SM} = c_{\gamma, SM} = 0 \, .
\label{smctree}
\eeq
Loop diagrams with the top quark and/or the $W$ induce a coupling to gluons and photons, which can be encoded as a contribution to the coefficients $c_g$ and $c_\gamma$ respectively.
In the SM they are given by,%
\beq
\hat c_{g,SM} 
= \frac{1}{2} F_{1/2}(x_t), \quad 
\hat c_{\gamma,SM} 
= 3 \left( \frac{2}{3} \right)^2 F_{1/2}(x_t) - F_1(x_W) \, ,
\label{smcloop}
\eeq
where $x_i = 4 m_i^2/m_h^2$ and the functions $F_{1/2,1}$ are given by,
\beq
F_{1/2}(x) &=&  2 x [1+(1-x) f(x)] \, ,\\
F_{1}(x) &=& 2 + 3 x + 3 x (2-x) f(x) \, , 
\eeq
where $f(x) = [\sin^{-1}(1/\sqrt{x})]^2$ for $x\geqslant1$ as it is the case for the top and $W$, since $m_h \simeq 125 \GeV$. 
The numerical values of the coefficients are then $\hat c_{g,SM} \simeq 2/3$ and $\hat c_{\gamma,SM} \simeq -6.5$.
These values are very close to the prediction from the higgs low-energy theorems, that is $\hat c_{g} = -(b^{(3)}_{t_R} + b^{(3)}_{t_L}) = 2/3$ and $\hat c_{\gamma} = -(b^{(EM)}_{W_T^\pm} + b^{(EM)}_{H} + N_c b^{(EM)}_{t_R} + N_c b^{(EM)}_{t_L}) \simeq -6.1$,
where we have made explicit the different contributions from each chirality and have separated the contributions from transverse and longitudinal components of the $W^\pm$.

As reviewed in Section~\ref{sec:pheno}, the couplings of the dilaton depart from those of the SM higgs.
Such modifications, when encoded in the parameters of the Lagrangian \eq{Leff}, read,
\beq
c_{V, \chi} = \frac{v}{f} \, ,
\eeq
\beq
c_{t, \chi} = \frac{v}{f} (1+\gamma_t) \, , \,\,\, 
c_{b, \chi} = \frac{v}{f} (1+\gamma_b) \, , \,\,\, 
c_{\tau, \chi} = \frac{v}{f} (1+\gamma_\tau) \, ,
\eeq
\beq
c_{g,\chi} = \frac{v}{f} (b_{IR}^{(3)}-b_{UV}^{(3)}) \, , \,\,\, 
c_{\gamma,\chi} = \frac{v}{f} (b_{IR}^{(EM)}-b_{UV}^{(EM)}) \, .
\label{cdilaton}
\eeq
All the coefficients are suppressed by the ratio $v/f$.
Further, the couplings to fermions depend on the anomalous dimension of the associated Yukawa coupling, while the coupling to massless gauge bosons depends on the strong sector's contribution to the corresponding $\beta$ function.
Recall that $b^{(\mathcal{J})}_{IR}$ includes the light composites only, specifically the NGB's of electroweak symmetry breaking (the charged components of the SM higgs doublet), and, although more model dependent, the right-handed top quark.
Therefore $b^{(3)}_{IR} = b^{(3)}_{t_R} = -1/3$ and $b^{(EM)}_{IR} = N_c b^{(EM)}_{t_R} + b^{(EM)}_{W^\pm_L} = - 11/9$.\\

When the coefficients of \eq{Leff} depart from the SM, decay rates and production cross sections change.
To first approximation, one can simply write for the former,%
\beq
\frac{\Gamma_{WW}}{\Gamma_{WW,SM}} = \frac{\Gamma_{ZZ}}{\Gamma_{ZZ,SM}} \simeq |c_V|^2 \, , \,\,\, 
\frac{\Gamma_{bb}}{\Gamma_{bb,SM}} \simeq |c_b|^2 \, , \,\,\, 
\frac{\Gamma_{\tau \tau}}{\Gamma_{\tau \tau,SM}} \simeq |c_\tau|^2 \, .
\eeq 
The decay widths to gluons and photons have a more complicated expression because of the interplay between the direct contribution from $c_g$ and $c_\gamma$, and the modification to the couplings of the particles running in the loop.
One can write
\beq
\frac{\Gamma_{gg}}{\Gamma_{gg,SM}} \simeq \frac{|\hat{c}_g|^2}{|\hat{c}_{g,SM}|^2} \, , \,\,\, \frac{\Gamma_{\gamma \gamma}}{\Gamma_{\gamma\gamma,SM}} \simeq \frac{|\hat c_\gamma|^2}{|\hat c_{\gamma,SM}|^2}
\eeq
where at the one-loop level
\beq
\hat{c}_g &=& c_g + c_t \frac{1}{2} F_{1/2}(x_t) \, , \\
\hat{c}_\gamma &=& c_\gamma + c_t \frac{4}{3} F_{1/2}(x_t) - c_V F_1(x_W) \, .
\eeq
As explained in Section~\ref{sec:traceanomaly}, those composite SM particles ($W^\pm_L$ and $t_R$) which get large masses $m > m_\chi/2$ (in practice all of them), partly decouple in their contribution to the dilaton coupling to massless gauge bosons,
due to the relation between the trace anomaly contribution and the triangle loop diagrams.
These then read
\beq
\hat c_{g,\chi} &\simeq& \frac{v}{f} \left( b_{IR}^{(3)} - b_{UV}^{(3)} + \frac{1}{2} F_{1/2}(x_t) \right) \equiv \frac{v}{f} b_{eff}^{(3)} \, , \\
\hat c_{\gamma,\chi} &\simeq& \frac{v}{f} \left( b_{IR}^{(EM)} - b_{UV}^{(EM)} + \frac{4}{3} F_{1/2}(x_t) - F_1(x_W) \right) \equiv \frac{v}{f} b_{eff}^{(EM)}\, ,
\eeq
where we have assumed that $c_{t,\chi} \approx v/f$.

The generic predictions for the dilaton are then an overall suppression of all decay rates by $v^2/f^2$, which will then be required to be close to one,
as expected by naturalness and electroweak precision tests (EWPT) (see Section~\ref{sec:constraints}).
The latter being so, enhancement of some of the decay rates is plausible, if large anomalous dimensions are present, for instance in the couplings to gluons and photons.
An extra suppression of the coupling to fermions by $\gamma < 0$ is not unplausible. 
\\

The dilaton production cross sections will also differ from those of the SM higgs.
At the Tevatron and LHC, the relevant production channels are gluon fusion (GF), vector boson fusion (VBF), and associated production with an electroweak vector boson (Vh).
One can express such cross section as,
\beq
\frac{\sigma_{GF}}{\sigma_{GF,SM}} \simeq \frac{|\hat{c}_g|^2}{|\hat{c}_{g,SM}|^2} \, , \,\,\,
\frac{\sigma_{VBF}}{\sigma_{VBF,SM}} \simeq |c_V|^2 \, , \,\,\,
\frac{\sigma_{Vh}}{\sigma_{Vh,SM}} \simeq |c_V|^2 \, .
\eeq
Therefore, for the dilaton one can expect a reduction in any of the production channels, unless the coupling to gluons is enhanced by a large $b_{UV}^{(3)}$, in which case the gluon fusion process could be larger than in SM.

\subsection{Constraints from EWPT and LHC data}
\label{sec:constraints}


Previous to the recent discovery at the LHC, indirect contraints on the higgs couplings, in particular $c_V$, were coming from EWPT.
These arise from the higgs one-loop contribution to the vector boson self energies.
When compared to the SM prediction, the additional contributions due to $c_V \neq 1$ to the parameters ${\hat T}, {\hat S}$ \cite{Barbieri:2007bh} is%
\beq
\Delta \hat T = - \frac{3 \alpha}{16 \pi \cos^2 \theta_W} (1-c_V^2) \log \left( \frac{\Lambda^2}{m_h^2} \right)\, , \,\,\, 
\Delta \hat S= + \frac{\alpha}{48 \pi \sin^2 \theta_W} (1-c_V^2) \log \left( \frac{\Lambda^2}{m_h^2} \right)\, ,
\eeq
where we assume that the logarithmically divergent one-loop contribution is cut at $\Lambda$. 
For our dilaton scenario one expects $\Lambda \simeq 4 \pi f = 4 \pi v / c_{V, \chi}$.
The one parameter fit, for $m_h = 125 \GeV$, yields the 99\% CL allowed region $ 0.86 \lesssim c_V^2 \lesssim 1.41$ \cite{Barbieri:2004qk} and thus the constraint $v/f \geqslant 0.93$.
One must keep in mind that this bound is obtained under the assumption of no extra UV contributions to ${\hat T}$ and ${\hat S}$.
While a tree-level $\hat{T}_{UV}$ can be forbidden by invoking custodial symmetry, one typically expects tree-level contributions coming from (\ref{Amix}) to $\hat{S}_{UV}$ of order $m_W^2/\Lambda^2 \sim 7 \times 10^{-4} (v^2/f^2)$.
\\

\begin{figure}[!t]
\begin{center}
\includegraphics[width=3in]{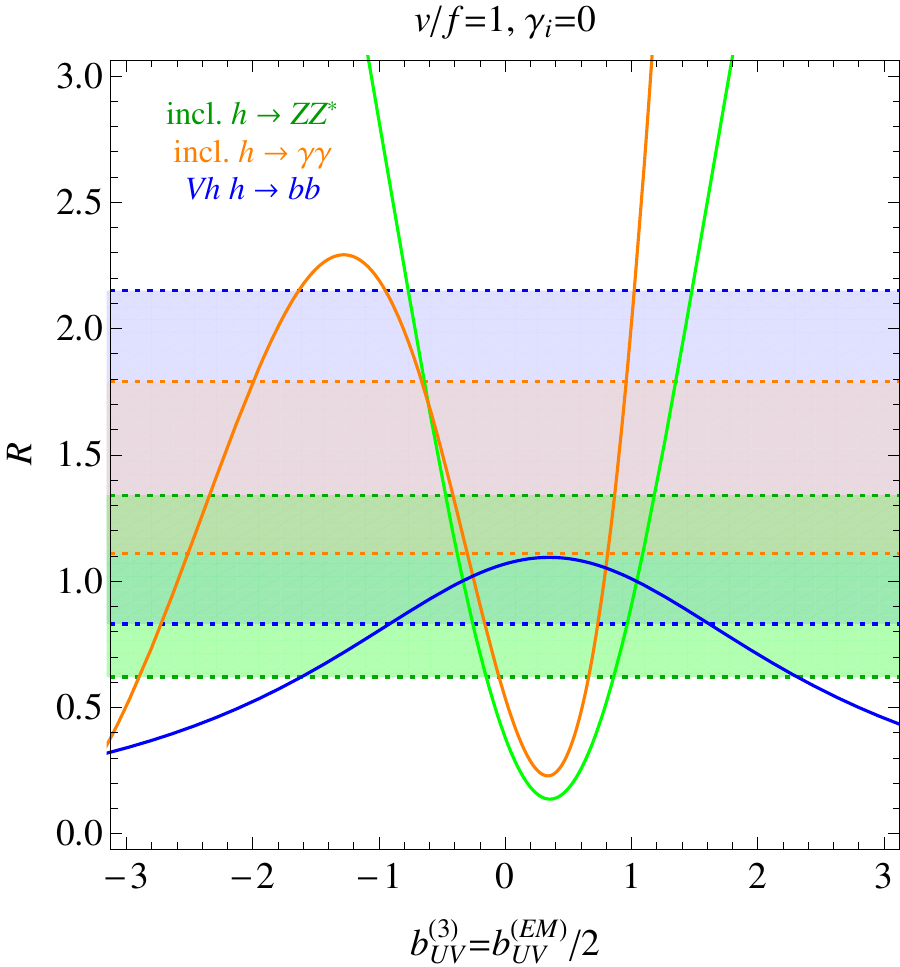}
\includegraphics[width=3in]{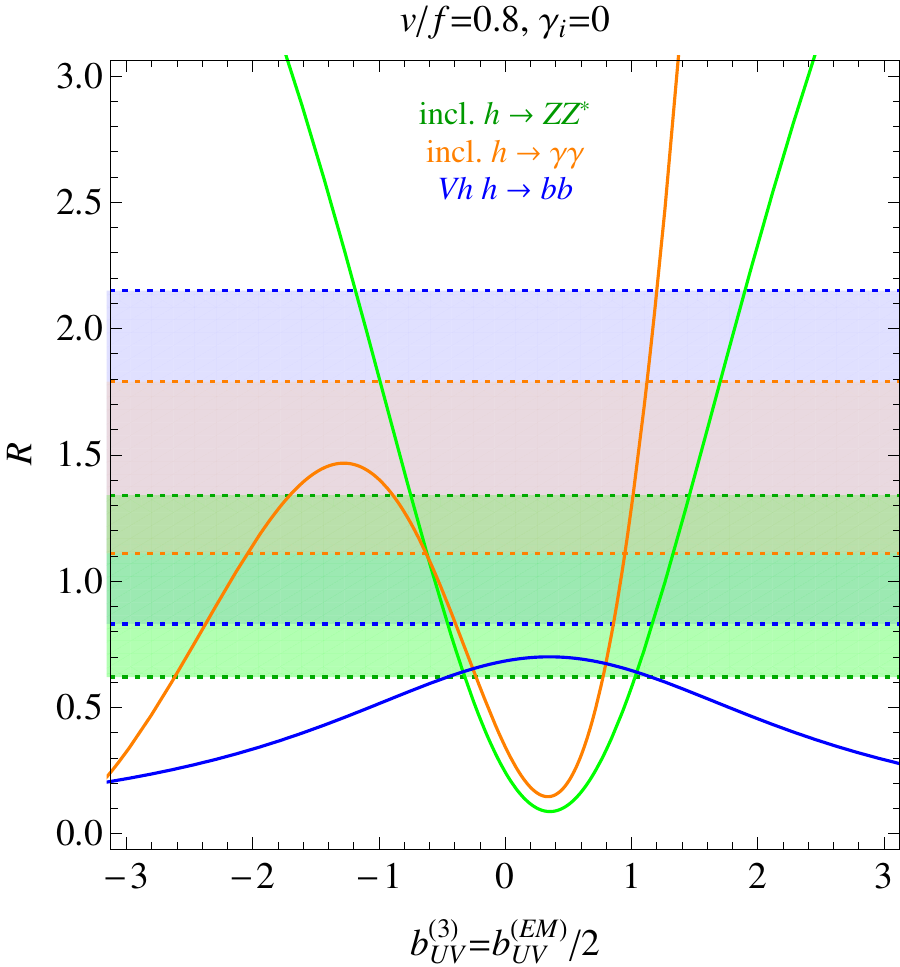}
\caption{Dilaton predictions for the rates $R_{incl., ZZ}$ (green line), $R_{incl., \gamma \gamma}$ (orange), and $R_{VH,bb}$ (blue) as a function of $b_{UV,CFT}^{(3)} = b_{UV,CFT}^{(EM)}/2$ for $v/f = 1$ (left panel) and $v/f = 0.8$ (right). Also shown as horizontal bands the current experimental intervals at $1\sigma$ CL (same color code).}
\label{Rs}
\end{center}
\end{figure}

Decay rates and production cross sections are the necessary ingredients to compare with Tevatron and LHC higgs data.
This is given in terms of the rates of each individual channel $j \to i$ (or combinations of) normalized to the SM prediction, 
\beq
R_{ji} \equiv [\sigma_{j \to h} \times \BR_{h \to i}]/[\sigma_{j \to h} \times \BR_{h \to i}]_{SM} \,. 
\label{R}
\eeq
There is already an extensive literature on constraints for the coefficients in \eq{Leff} obtained by fitting the $R_{ij}$'s.
The current errors on these are large, however strong correlations among the actual multi-dimensional fit parameters are obscured if one considers only the limits on individual coefficients. 
For this reason, in this section we directly compare the results of our theoretical predictions with the experimental values of the rates \cite{Carmi:2012in}. 
It is useful to present the scaling of the different $R_{ij}$ with the dilaton parameters, that is $v/f$ and the anomalous dimensions $\gamma_i, b_{eff}^{(\mathcal{J})}$.
The total decay rate of the dilaton compared to the SM can be approximated (if the deviations of the couplings are small) by
\begin{figure}[!t]
\begin{center}
\includegraphics[width=3in]{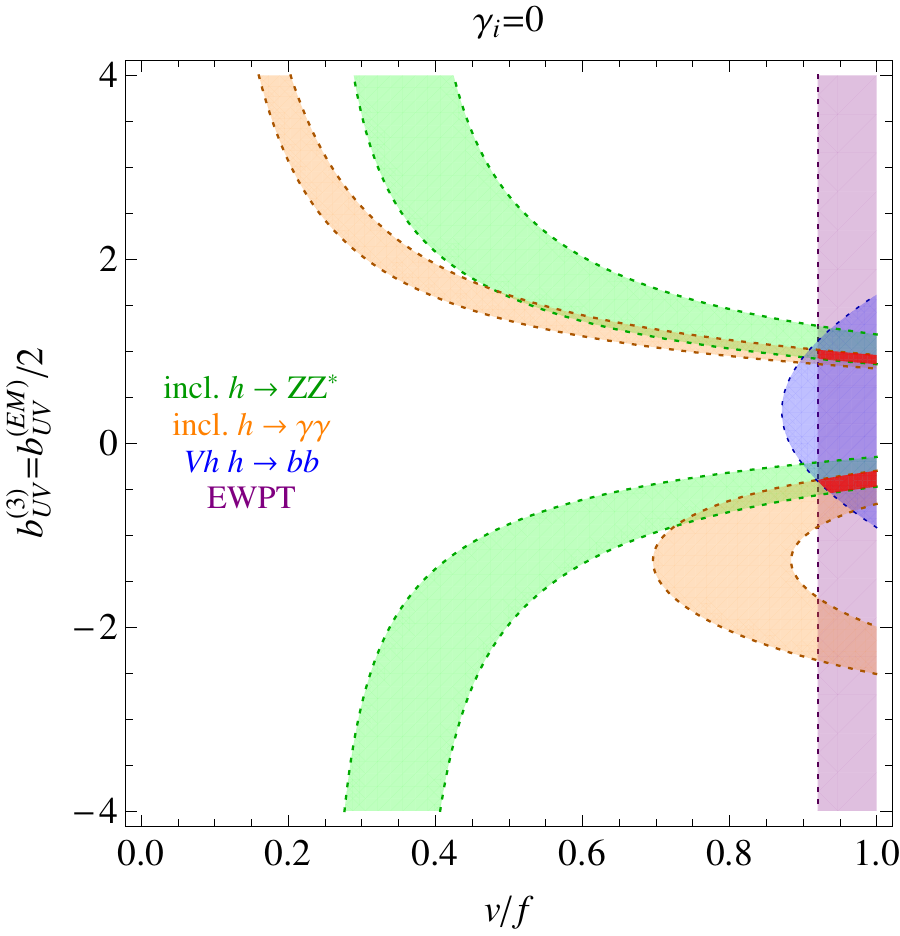}
\includegraphics[width=3in]{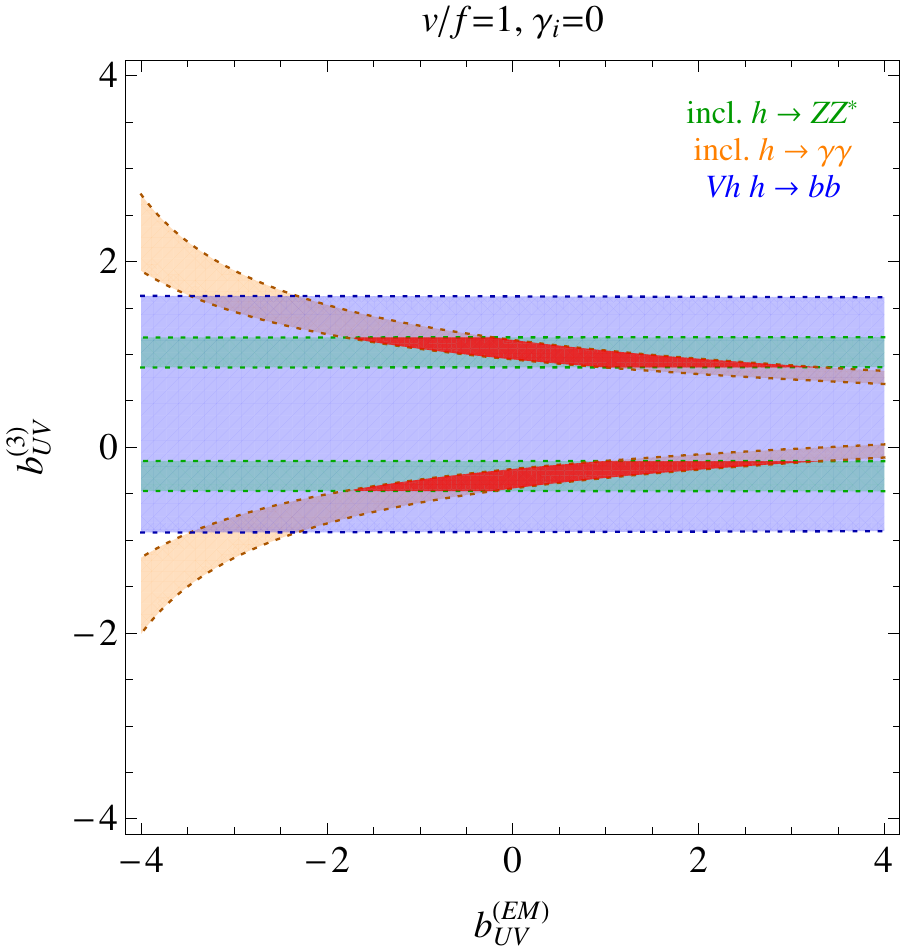}
\caption{Left: Constraints on the $v/f$ and $b_{UV,CFT}^{(3)} = b_{UV,CFT}^{(EM)}/2$ plane (shaded allowed regions) from experimental measurement at the $1\sigma$ CL of the rates $R_{incl., ZZ}$ (green), $R_{incl., \gamma \gamma}$ (orange), $R_{VH,bb}$ (blue), and EWPT at 99\% CL (purple). 
The overlap region is shown in red.
We have assumed $\gamma_i = 0$, and we recall that $c_{V, \sigma} = v/f$.
Right: Same constraints in the $b_{UV,CFT}^{(EM)}$ and $b_{UV,CFT}^{(3)}$ plane fixing $v/f = 1$.}
\label{bchi}
\end{center}
\end{figure}
\beq
|C_{tot}|^2 &=& \frac{\Gamma_{tot,\chi}}{\Gamma_{tot,SM}} \simeq \frac{v^2}{f^2}
\left[ \BR_{WW,SM} + \BR_{ZZ,SM} + (1+\gamma_b) \BR_{bb,SM} + \frac{(b_{eff}^{(3)})^2}{(b_{t}^{(3)})^2} \BR_{gg,SM} \right] \nonumber \\ 
&\equiv& \frac{v^2}{f^2} C^2\, .
\eeq
With this we can compute the rates as $R \simeq (\sigma \Gamma) / (\sigma \Gamma)_{SM} \times |C_{tot}|^{-2}$, and one obtains for the individual production channels,
\beq
&&R_{GF,(WW,ZZ)} \simeq \frac{v^2}{f^2} \frac{1}{C^2} \left(\frac{{b_{eff}^{(3)}}}{{b_t^{(3)}}}\right)^2   \, , \,\,\,
R_{GF,\gamma\gamma} \simeq \frac{v^2}{f^2} \frac{1}{C^2} \left(\frac{ b_{eff}^{(3)} \, b_{eff}^{(EM)} }{b_{t}^{(3)} \, b_{t+W}^{(EM)}}\right)^2   \, , \,\,\,
\nonumber \\
&& R_{GF,\tau \tau} \simeq \frac{v^2}{f^2} \frac{1}{C^2} \left( \frac{b_{eff}^{(3)} \, (1+\gamma_\tau)}{b_t^{(3)}} \right)^2   \, , \,\,\, 
R_{VBF,\gamma\gamma} \simeq \frac{v^2}{f^2} \frac{1}{C^2} \left( \frac{b_{eff}^{(EM)}}{b_{t+W}^{(EM)}}\right)^2 \, , \,\,\, \nonumber \\
&& R_{VBF,(WW,ZZ)} \simeq \frac{v^2}{f^2} \frac{1}{C^2} \, , \,\,\,  
R_{VBF,\tau \tau} \simeq \frac{v^2}{f^2} \frac{1}{C^2} (1+\gamma_\tau)^2 \, , \,\,\,
R_{Vh,bb} \simeq \frac{v^2}{f^2} \frac{1}{C^2} (1+\gamma_b)^2 \, . \nonumber \\
\eeq
All the rates scale as $v^2/f^2$, and the inclusive modes as well, since all coefficients in \eq{Leff} for the dilaton are proportional to $v/f$, and likewise for $|C_{tot}|$.
Paying attention to the individual channels one can gain information on the anomalous dimensions.
We show in \fig{bchi} the constraints from the present measurements of three different rates: inclusive higgs production and decay to $ZZ$ or to $\gamma \gamma$, $R_{incl., ZZ}$ and $R_{incl., \gamma\gamma}$ respectively, and associated vector boson production and decay to $b \bar b$, $R_{Vh, bb}$.
From the left panel one can see the preference of the data for values of $v/f$ very close to one, as was already suggested by EWPT (also shown as a vertical strip).
This is driven by the measurement of $R_{VH,bb}$, since we assumed no deviations in the coupling to the bottom except for the $v/f$ factor.
The inclusive measurements $R_{incl., ZZ}$ and $R_{incl., \gamma \gamma}$ are instead sensitive to the coefficients of the $\beta$ function.
In particular, as shown in the right panel of \fig{bchi}, $R_{incl., ZZ}$ delimits the preferred values for $b^{(3)}_{UV}$, while the overlap with $R_{incl., \gamma \gamma}$ does this for $b^{(EM)}_{UV}$.
We also show in \fig{Rs} the prediction for these three rates as a function of $b^{(3)}_{UV} = b^{(EM)}_{UV}/2$ (this choice correspond to the symmetric scenario $b^{(1)}_{UV} = b^{(2)}_{UV} = b^{(3)}_{UV}$ as in \cite{Frigerio:2011zg}), and its overlap with current measurements at $1\sigma$ CL.
Enhancement of the $ZZ$ and $\gamma \gamma$ rates are easily obtained for both $v/f =1$ (left panel) and $v/f = 0.8$ (right panel).
The difference between negative and positive values of $b^{(3)}_{UV}$ is due to the difference in sign of the SM contribution to $\hat c_g$ and $\hat c_\gamma$.
Finally, notice that the $b{\bar b}$ rate from associated production is generically suppressed, due to the lack of enhancement in the production cross section.
This conclusion would not be changed by turning on $\gamma_b \neq 0$, since the $b{\bar b}$ channel already dominates the decay of the higgs for $\gamma_b = 0$.

\section{General considerations for the dilaton mass}
\label{sec:general}
\setcounter{equation}{0}

The main difference between a standard Goldstone boson arising from an internal global symmetry and the dilaton is that scale invariance allows for a non-derivative quartic self coupling,
which plays a crucial role in the discussion of the SBSI:
\begin{equation}
S=\int d^4x \frac{f^2}{2}(\partial\chi)^2-a f^4\chi^4 +\mbox{higher derivatives}
\label{eq:effpotential}
\end{equation}
The presence of this term will make it very difficult to achieve the SBSI. When $a\neq 0$ the theory is either forced to $f\to \infty$ for $a<0$ (a runaway direction), or to $f=0$ for $a>0$. Thus one needs to tune $a=0$ in the effective theory (as explained by Fubini~\cite{Fubini:1976jm}). In order to achieve SBSI one needs to relax 
$a=0$ to $|a|\ll1$, so that the broken phase $\langle\chi\rangle=1$ is only metastable. 
Adding an explicit breaking term to the CFT with an almost marginal operator
\begin{equation}
\delta S=\int d^4 x \lambda(\mu)\mathcal{O}
\end{equation}
gives rise, in general, to an effective potential for the dilaton of the form
\begin{equation}
V(\chi)=\chi^4  F(\lambda(\chi)) \, , 
\label{eq:dilatonpotential}
\end{equation}
where $F$ is a function of $\lambda$ which parametrizes the explicit breaking of scale invariance as a non-trivial function of $\chi$.
This potential is of the Coleman-Weinberg type when $\lambda$ is almost marginal. 
Then, as
explained by Weinberg~\cite{Weinberg} and also stressed by Rattazzi and Zaffaroni~\cite{RattazziZaffaroni}, a
natural SBSI along with the generation of a large hierarchy of scales is possible within naturalness. For this one needs $a$ to be small (as assumed) and $\mathcal{O}$ to be  a marginally relevant deformation (as in QCD) while $\lambda$ remains perturbative over the relevant range of renormalization group running. 
In this case $F(\lambda(f))$ can have a minimum at a scale $f\gg \Lambda_s$, where $\Lambda_s$ is the scale where $\lambda$ would become non-perturbative. Because $f\gg\Lambda_s$, $\lambda$ stays perturbative and the dilaton remains light, that is scale invariance can be spontaneously broken. 
The stationary condition of $V$ is
\begin{equation}
\label{minimum}
V^\prime=f^3\left[4 F(\lambda(f))+\beta F^\prime(\lambda(f))\right]=0
\end{equation}
which results in a dilaton mass
\begin{equation}
\label{dilaMass}
m^2_{dil}=f^2\beta \left[\beta F^{\prime\prime}+4 F^\prime+ \beta^\prime F^\prime \right]\simeq 4 f^2\beta F^\prime(\lambda(f))=-16f^2 F(\lambda(f))
\end{equation}
where $\beta^\prime=d\beta/d\lambda$. In the second equality we have also assumed that $\beta^\prime\ll 1$.
An explicit (supersymmetric) example illustrating how this mechanism can work will be presented in the next section. The Goldberger-Wise stabilization mechanism for the RSI model is also an example for this mechanism, as we will discuss in detail in Section~\ref{sec:GW}.

The main questions related to the naturalness of this mechanism are then why is $F\ll 1$ at the minimum (or, for a perturbative expansion in $\lambda$, $a\ll 1$) along with $\beta \ll 1$, and why are we allowing only almost marginal perturbations. Let us start with $F\ll 1$. The case $F=0$ corresponds to a situation with no potential for the dilaton, and thus an arbitrary value of $f$ is allowed. This means that there is a flat direction in the theory. The presence of flat directions is quite natural in supersymmetric theories, however no non-supersymmetric example of physically inequivalent flat directions is known.\footnote{The only other known way of generating flat directions is via the Goldstone theorem, but that will not generate physically inequivalent vacua as is required for the case with an arbitrary scale $f$.} The closest anyone has been able to get to this situation were the so-called orbifold gauge theories obtained via projecting out some of the fields and couplings of an ${\cal N}=4$ SUSY gauge theory~\cite{BershadskyVafa}. In this case the large-$N$ limit of the $\beta$ functions agrees with those of the SUSY theories, however $1/N$ corrections lift the flat directions~\cite{CST}. 

The other question is why only close-to-marginal perturbations are allowed, as these are the only ones that would allow for a light dilaton. 
This part of the naturalness problem is thus rephrased in terms of what relevant deformations the CFT supports. If it turns out that only marginal perturbations are possible then a light dilaton is a natural possibility (once the flat direction is present). 
Do such theories exist? Again, SUSY theories (SCFT's), especially chiral ones, give a handle on this because of the non-renormalization theorem: the relevant deformations (if there are any) can be made naturally small. For the case of non-supersymmetric CFT's one would expect that only chiral gauge theories might have a chance of  giving a naturally light dilaton, but even those face the question of the origin of a flat direction.

Let's try to estimate how much fine tuning is hidden in these assumptions. 
The minimization condition (\ref{minimum}) says that for $\beta\ll 1$ the quartic $F$ must  almost vanish. In turn this ensures that the dilaton mass (\ref{dilaMass}) 
can be made parametrically smaller than $f$. In other words, if we start with an almost flat direction, $F\ll 1$, then we can easily stabilize it by a small breaking controlled by $\beta$. However, the starting assumption of almost flatness is itself plagued by 
fine-tuning unless a symmetry reason can be invoked. In fact, the NDA for the quartic is
\begin{equation}
F_{NDA}\sim \frac{\Lambda^4}{16\pi^2 f^4}\sim 16\pi^2
\end{equation}
making the minimization condition (\ref{minimum}) behind the flatness of the potential and the lightness of the dilaton very unlikely to be realized in a generic theory. With such a large quartic the dilaton mass would  be at the cutoff $m_{dil}^2 \sim \Lambda^2$,
and the explicit breaking of scale invariance necessarily large, 
\beq
\beta \sim \frac{4 F_{NDA}}{F'_{NDA}} \sim 4 \pi.
\eeq
As we explain in more detail below, this is the situation realized in QCD-like or technicolor theories, where the gauge coupling $g^2$, to be identified with $\lambda$, becomes non-perturbative. No light scalar degree of freedom with the properties of the dilaton is expected to be present in the spectrum. 

The above naive estimates can be refined for theories where the explicit breaking of scale invariance comes from a coupling external to the strong conformal sector.
In general its $\beta$ function will be given by
\beq
\label{betaperturb}
\beta(\lambda)= \frac{d \lambda}{d \ln \mu} = \epsilon \lambda + \frac{b_1}{4\pi} \lambda^2 + \order{\lambda^3}
\eeq
which is under control (\ie small) as long as $\lambda$ remains perturbative, $\lambda \lesssim 4 \pi$, for $b_n \sim \order{1}$, ($\epsilon = b_0$). Here $\epsilon$ is identified as the deviation from marginality of the perturbing operator, $|\epsilon| < 1$, which is set by the strongly coupled CFT.  The perturbativity of $\lambda$  is a necessary condition to obtain a parametrically light dilaton, unless one is willing to accept that even in the non-perturbative regime, the $\beta$ function remains small but non-zero over a large range of values of the coupling constant, which is a very special dynamical assumption, and we know of no 
examples of such theories.

The consistency of a perturbative expansion in $\lambda$ with the requirement of SBSI and the generation of a large hierarchy is determined by the minimization condition (\ref{minimum}), and can only be achieved by reducing the intrinsic dilaton
quartic $a$ to values comparable with the symmetry breaking contributions
\beq
F(\lambda)=(4\pi)^2\left[c_0+\sum_n c_n \left(\frac{\lambda}{4\pi}\right)^n \right]  \, , \quad c_0\ll c_n \sim 1\,,\quad  a=(4\pi)^2 c_0 \, .
\eeq
Then the minimization condition (\ref{minimum}), expanded in powers of $\lambda$ and $\epsilon$, yields $\lambda(f) \simeq 4 \pi c_0/c_1 \simeq 4\pi/\Delta$, where $\Delta$ is the amount of fine tuning. The coupling $\lambda$ is allowed to remain perturbative at the minimum.
From the dilaton mass formula (\ref{dilaMass})
\begin{equation}
\frac{m_{dil}^2}{\Lambda^2} \sim \frac{\beta}{\pi}\simeq \epsilon \frac{\lambda}{\pi}
\label{dilmassexpl}
\end{equation}
then we understand that the fine tuning is tied to the dilaton mass, and further that it is moderated by the marginality of the symmetry breaking coupling
\begin{equation}
\Delta=\frac{4\pi}{\lambda} \sim 4\epsilon \frac{\Lambda^2}{m_{dil}^2}\,.
\end{equation}
From this formula it appears that the fine-tuning can be reduced arbitrarily by taking $\epsilon\to 0$, however one should not forget that once $\epsilon$ is very small the next-to-leading term in (\ref{betaperturb}) will become the dominant source of the dilaton mass, replacing (\ref{dilmassexpl}) with 
\begin{equation}
\frac{m_{dil}^2}{\Lambda^2} \sim \frac{\beta}{\pi} \sim   \frac{\lambda^2}{4\pi^2}
\label{dilmassmin}
\end{equation}
 so that $\Delta$ scales, at best, linearly\footnote{We assumed that the leading symmetry breaking term $c_1$ is not suppressed. In case instead $c_2$ term dominates over $c_1$ the fine tuning scales as $\Delta=(4\pi/\lambda)^2\gtrsim (2\Lambda/m_{dil})^{4/3}$.} with $\Lambda/m_{dil}$
\begin{equation}
\Delta\gtrsim 2\Lambda/m_{dil} \simeq 50 \left(\frac{f}{246\mathrm{ GeV}}\right)\left(\frac{125\mathrm{ GeV}}{m_{dil}}\right)\,.
\end{equation}
From this discussion, in particular \eq{dilmassmin}, one can again see that in technicolor theories, where $\epsilon = 0$ and  $\lambda = g^2$ is required to become non-perturbative to generate a condensate, $m_{dil} \sim \Lambda$.

Finally notice that if we define the perturbative coupling $\lambda$ at some scale $M$ where the strongly coupled theory is conformal then a large hierarchy of scales $f \ll M$ is generated because of the assumption that $\mathcal{O}$ is almost marginal
\beq
f \simeq M \left(\frac{-4\pi c_0}{\lambda(M) c_1} \right)^{1/\epsilon}~.
\eeq

In the next section we present a natural supersymmetric implementation of the mechanism outlined above for a naturally light dilaton. We will see that SUSY will ensure the presence of a flat direction, which will be slightly broken by a non-perturbative effect, giving rise to a runaway direction $a<0, |a|\ll 1$, which will then be stabilized by a small, technically natural almost-marginal deformation at $f\gg \Lambda_s$. SUSY plays a crucial role in all aspects of the naturalness of the light dilaton in this model.

\section{The 3-2 model: an illustrative SUSY example}
\label{sec:3-2}
\setcounter{equation}{0}

A simple model that illustrates the general discussion related to the magnitude of the dilaton mass is the well-known 3-2 model of dynamical supersymmetry breaking~\cite{Affleck:1985xz}. The model is given by the following chiral superfield matter content under an $SU(3) \times SU(2)$ ${\cal N}=1$ supersymmetric gauge theory:

\beq
\begin{array}{c|cc|cc}
 & SU(3) & SU(2) & U(1) & U(1)_R\\ \hline
Q & \Yfund & \Yfund & 1/3 & 1\\
L &  {\bf 1} &  \Yfund & -1 & -3 \\
\overline{U}  &  \overline{\Yfund} &  {\bf 1} & -4/3 & -8 \ \\
\overline{D} & \overline{\Yfund}  &  {\bf 1} &2/3 & 4 
\end{array}~.
\eeq
together with a tree-level superpotential 
\beq
 W =  
\lambda \, Q \bar{D} L~,
\label{eq:Wtree}
\eeq
This theory is an ideal toy example because in the $\lambda\to 0$ limit the theory has classical flat directions that are parametrized by the invariants $Q\bar{D}L$, $Q\bar{U}L$ and det$(\bar{Q}Q)$, where $\bar{Q}=(\bar{U},\bar{D})$. All of these flat directions are lifted by the addition of the superpotential. However in the limit when $\lambda \ll 1$ this potential will be very shallow. In the limit when the $SU(3)$ group is much stronger than the SU(2) group, $\Lambda_3\gg \Lambda_2$, the largest dynamical effect will be the presence of $SU(3)$ instantons generating a dynamical Affleck-Dine-Seiberg (ADS) superpotential of the form 
\beq
W_{\rm dyn} = {{\Lambda_3^{7}}\over{{\rm det} (\overline{Q} Q) }} ~,
\label{eq:ADSsuppot}
\eeq
This superpotential will  force the fields to large expectation values, and without the stabilizing tree-level superpotential term in (\ref{eq:Wtree}) one would have a runaway direction. For $\lambda \ll 1$ the stabilized field values will be $\gg \Lambda_3$, and the gauge group will be completely broken. Thus for sufficiently small $\lambda$ the gauge symmetry will be broken dynamically via the instanton effects before the gauge group becomes strongly coupled. The theory is approximately conformal, only broken by the weak gauge couplings and the very weak $\lambda$. This implies that there is also a dynamical spontaneous breaking of the approximate conformal symmetry, and that one expects a light dilaton field as long as the field VEVs satisfy 
$f=\langle \Phi \rangle \gg \Lambda_3$. 

Since the theory is calculable for $\lambda \ll 1$ one can explicitly verify this. The crude estimate for the dilaton mass assumes that all field values are roughly of the same order $\langle \phi \rangle \sim f$ with $f \gg \Lambda_3$ for $\lambda \ll 1$. In this case the potential is of the order 
\beq
V \approx \frac{\Lambda_3^{14}}{f^{10}} 
+ \lambda\frac{\Lambda_3^{7}}{f^{3}} + \lambda^2 f^4~,
\label{eq:potential}
\eeq
Minimizing this potential one obtains the scaling of the VEV and of the vacuum energy:
\beq
f \approx \frac{\Lambda_3} {\lambda^{{1}/{7}}}~, \ \ V \approx \lambda^{{10}/{7}} \Lambda_3^4.
\eeq
Thus using the usual parametrization $ \phi = f e^{\sigma/f}$ we find that the dilaton mass is of order 
\begin{equation}
m_{dil} \approx \lambda f \approx \lambda^{\frac{6}{7}} \Lambda_3 
\label{eq:32dilmass}
\end{equation}
in agreement with \cite{Bagger:1994hh}.
Thus we obtain a naturally light dilaton, below the scale of conformality breaking $f$
as long as $\lambda \ll 1$ (and also below the dynamical scale $\Lambda_3$).
This is because the theory is weakly coupled in that regime, and one has a predominantly  spontaneous breaking of the conformal symmetry.  The main source of the dilaton mass here is the instanton effect itself, which is not scale invariant. One can easily check that loop corrections due to the running of $\lambda$ and the gauge coupling $g_3$ result in subdominant corrections. 

Once $\lambda \sim {\cal O}(1)$ the theory enters the strong coupling regime before conformality is broken. In this case there is a large explicit breaking of scale invariance both due to the running of the coupling and the instantons in the strongly coupled group. The dilaton mass is no longer expected to be suppressed compared to $\Lambda_3$ as suggested by (\ref{eq:32dilmass}), even though the actual expression for the dilaton mass is no longer calculable. While the superpotential is still exact, the K\"ahler potential will get large corrections and even the right degrees of freedom may change. Nevertheless one does not expect these effects to provide any suppression of the dilaton mass. In particular (\ref{eq:32dilmass}) suggests that for $\lambda \sim 4 \pi$ we get $m_{dil}\sim 4\pi f$ around the cutoff scale of the effective theory.

The main lesson from this example is that having a light dynamical dilaton is possible, however it seems crucial to have a weakly coupled flat direction available in the theory. It is hard to imagine such flat directions without the presence of supersymmetry.

\section{The Goldberger-Wise stabilized Randall-Sundrum\\ model: a non-SUSY example}
\label{sec:GW}
\setcounter{equation}{0}

The most often discussed non-SUSY example for a model with SBSI is the RSI model~\cite{Randall:1999ee} with the GW stabilization mechanism \cite{Goldberger:1999uk}. Here a warped extra dimension with an AdS$_5$ background and curvature radius $R$ is cut off via a UV brane with tension $V_0$ and an IR brane with tension $V_1$. The location of the IR brane $R'$ (usually referred to as the radion~\cite{CHL,Goldberger:1999uk,CGK,CGRT,GW2}) provides the dilaton, with the identification $1/R'=f$. The GW stabilization~\cite{Goldberger:1999uk} adds a bulk scalar with a very small bulk mass $m$, such that $\epsilon \sim m^2 R^2/4\ll 1$. The standard lore is that this model is a non-SUSY example with a naturally light dilaton.   We are interested in the question whether the appearance of the light dilaton is indeed natural.  For this we need to identify the effective potential (\ref{eq:effpotential}) of the dilaton. This effective potential depends on the bulk and brane tensions and also on the parameters of the GW stabilizing field. 

The potential in the absence of the GW field is given by
\begin{equation}
V_{eff}= V_0 +V_1 \left( \frac{R}{R'}\right)^4 +\Lambda_{(5)} R \left(1-\left(\frac{R}{R'}\right)^4\right) \, ,
\end{equation}
where $\Lambda_{(5)}$ is the 5D cosmological constant. In terms of the dilaton $f =1/R'$ this is written as
\begin{equation}
V_{eff}(\chi )= V_0+\Lambda_{(5)} R +  f^4 \left( V_1R^4 -\Lambda_{(5)} R^5\right) \, .
\end{equation}
This potential is clearly of the form  (\ref{eq:effpotential}) as expected from SBSI, with an additional 4D cosmological constant. The usual RS tuning conditions consist of making the choice $V_0=-V_1 =-\Lambda_{(5)} R$. One of these tunings eliminates the 4D cosmological constant, while the second clearly corresponds to  tuning the coefficient of the quartic self-coupling $a$ to zero, that is to make the dilaton a flat direction. If we did not make this tuning, the two branes would  either collide or fly apart \cite{Csaki:1999jh}, depending on the sign of the quartic $V_1R^4-\Lambda_{(5)} R^5$. This is the origin of the tuning for the light dilaton mass in RS: the only reason for this second tuning is to ensure that the dilaton is flat in the absence of a stabilization mechanism, that is to ensure the spontaneous breaking of conformality.
Without this tuning of the IR brane tension there would be a large quartic.  This does not preclude the stabilization of the dilaton at large VEVs once the stabilizing, explicit breaking is introduced, it simply requires that the breaking is large (it can no longer be an almost marginal perturbation). In this case one expects a large dilaton mass of the order of the other scales in the theory, rather than a parametrically suppressed dilaton mass. 

In order to quantify the tuning let us give the NDA value for the size of this quartic $\delta a$, by estimating the bulk contribution $\Lambda_{(5)} R^5$ and the IR contribution $V_1 R^4$. To find the bulk contribution we need to first find the cutoff of a 5D gravitational theory $\Lambda^{grav}$ in terms of the 5D Planck scale $M_*$. NDA relates the two as $\Lambda^{grav}= M_* (24 \pi^3)^\frac{1}{3}$, where the $24 \pi^3$  is the 5D loop-factor~\cite{Kaustubh5DNDA}. The AdS curvature scale $R$ is given by 
$R^2=- 12 M_*^3/\Lambda_{(5)}$. NDA predicts the size of the 5D cosmological constant to be at least  $\Lambda_{(5)} \sim (\Lambda^{grav})^5/(24 \pi^3)$. Putting these all together one finds that the natural value of the bulk contribution to the tree-level quartic is of the order
\begin{equation}
\delta a_{(bulk)} \sim \Lambda_{(5)} R^5 \sim \frac{12^\frac{5}{2}}{24\pi^3} \sim {\cal O}(1).
\label{eq:deltaa}
\end{equation}
The IR contribution on the other hand is given by 
\begin{equation}
\delta a_{(IR)} = -V_1 R^4 = -V_1 \left(\frac{R}{R'}\right)^4 R'^4 =
 \frac{\tilde{V}_1}{\left(\frac{\Lambda}{4\pi}\right)^4} \, ,
\end{equation}
where $\tilde{V}_1$ is the warped-down physical value of the IR brane tension, and $\Lambda$ is the local cutoff of order $ 4\pi/R'$.  One can clearly see that the correction to this IR piece exactly matches the estimate for the 4D NDA value of the dilaton quartic: the one-loop correction to $\tilde V_1$ is just ${\Lambda}^4/(16\pi^2)$, yielding 
$\delta a_{(IR)}\sim 16\pi^2$. The tuning is then due to the fact that $\delta a = \delta a_{(bulk)}+\delta a_{(IR)}$ where the first term is ${\cal O}(1)$, which needs to cancel against the term that is ${\cal O}(16\pi^2)$. 
 
The GW stabilization mechanism instead {\it assumes} the original RS tuning for the brane tensions with $\delta a=0$, that is it assumes that the unperturbed situation corresponds to the theory with a flat dilaton. 
At that point the possibility of a small dilaton mass follows from the discussion in Section~\ref{sec:general}. Without that assumption the dilaton mass would generically not be light (and the back-reaction of the metric not negligible). Once the RS tuning assumption is made a light dilaton can be produced.

In order to find the amount of tuning in RS-GW we need to review the full dilaton potential.  This was first calculated in~\cite{Goldberger:1999uk} examined in great detail in \cite{RattazziZaffaroni,CGRT,GW2} and is given, in the interesting region where $f \ll 1/R$, by
\beq
V = f^4 \left\{ (4+2 \epsilon) \left[ {v}_1 - {v}_0 \left( f R \right) ^\epsilon \right]^2 - \epsilon v_1^2 + \delta a+  \order{\epsilon^2} \right\}
= f^4 F(f ) \, ,
\label{GWpot}
\eeq
where ${v}_{0,1}$ are the UV and IR VEVs of the GW scalar field in units of the AdS curvature $R$, and $\delta a$ is the tree-level contribution to the quartic from the mistuning of the IR tension estimated above. The interpretation of this potential is that it is the result of a UV deformation
\beq
\delta \Lag = \lambda \mathcal{O} \, ,
\eeq
where the operator $\mathcal{O}$ has dimension $4+\epsilon$, and with ${v}_0$ being related to the UV value of the coupling $\lambda$, while $v_1$ being related to the VEV 
$\langle {\cal O} \rangle$. Thus $v_0$ sets the UV value of the source of explicit breaking of conformality. The latter is parametrized by $\epsilon$, the departure from marginality of $\mathcal{O}$; $v_1$ sets the amount of spontaneous breaking in the limit $v_0 \to 0$, and can be further interpreted as ${v}_1 \sim \delta \lambda(f)$, an IR threshold contribution.
The actual running of $\lambda$ is given in general by (\ref{betaperturb}), 
which is the actual measure of the explicit breaking of the conformal symmetry.
This explains  why in \eq{GWpot}, where all $b_i$ have been neglected, any non-trivial dependence on $\phi$ vanishes for $\epsilon = 0$: 
$\epsilon$ is the contribution of the CFT to the running of $\lambda$, and when $\epsilon\to 0$ there is no explicit breaking of the CFT, so the dilaton potential can only contain a quartic term.
Note then that the parameters ${v}_{0,1}$ do not automatically break the conformal symmetry.

The minimum of the potential \eq{GWpot} is at
\beq
f = \frac{1}{R} \left( \frac{{v}_1 + \sqrt{-\delta a/4}}{{v}_0} + O(\epsilon) \right)^{1/\epsilon} \, ,
\label{GWvev}
\eeq
which is exponentially smaller than $1/R$, yielding the necessary hierarchy for the RS model.
Notice that this minimum is the solution of \eq{minimum} with a small $\beta$ function $\beta \simeq \epsilon \lambda$, that is $F(\lambda(f)) = O(\epsilon)$.
The assumption that $\epsilon$ is small is a necessary ingredient in order to reproduce a large hierarchy of scales, but the fact that the full $\beta$ function (\ref{betaperturb}) remains small at the minimum  is an unnatural outcome, precisely because it requires the vanishing of the dilaton quartic coupling at the minimum without any symmetry argument or dynamical mechanism behind it.

We can now quantify the amount of tuning needed in RS in order to obtain the right hierarchy. In (\ref{GWvev}) we need the correction of the quartic to be not too different from $v_1$:
\begin{equation}
\sqrt{-\delta a/4} \lesssim v_1
\end{equation}
yielding a tuning of order
\begin{equation}
\Delta = \frac{a}{|\delta a|} \gtrsim \frac{4 \pi^2}{v_1^2}\ .
\end{equation}
For example for the canonical choice of $v_0=1,v_1=1/10, \epsilon =1/15$ one finds $\Delta \sim 4000$, a per mil level tuning. Without this tuning there would be no hierarchical minimum with a small back reaction.
That would be realized with $\epsilon < 0$ (a relevant deformation) and $\lambda$ becoming non-perturbative at low energies, but then no trace of a dilaton associated with SBSI would remain.%
\footnote{To avoid this tuning and keep a light dilaton in the spectrum, one would like to find an explanation for why in the limit $\epsilon \to 0$ one also finds  $\delta a \to 0$, as in the construction of Contino, Pomarol, Rattazzi~\cite{RiccardoTalk}.}

Once the hierarchy is established, one can try to see if the radion can be made to have properties similar to the Higgs. It turns out that the main obstacle is to obtain $f\approx v$. The point is that the kinetic term of the radion (in the normalization we have used so far) is very large, it is in fact enhanced~\cite{CGK,CGRT,GW2} by the factor $N^2 = 12 (M_* R)^3$, which by the requirement that the gravitational theory is calculable should be $N\gg 1$.  This leads to an enhancement for the expression of the physical value of the scale of SBSI for the RS model:
\begin{equation}
f^{(RS)} =\frac{1}{R'} \sqrt{12 (M_* R)^3}
\end{equation}
This is the scale that will suppress all the dilaton couplings. If one were to do away with the Higgs doublet localized on the IR brane and try to substitute the radion for the higgs, the expression for $v/f$ using the basic relations for higgsless models~\cite{higgsless} would be given by 
\begin{equation}
\frac{v}{f^{(RS)}} = \frac{2}{g} \frac{1}{N\sqrt{\log \frac{R'}{R}}}\ .
\end{equation}
For calculable gravity models with a hierarchy one finds $v/f\ll 1$. Alternatively one could consider a theory with a very heavy higgs on the IR brane, in which case one just finds 
\begin{equation}
\frac{v}{f^{(RS)}} = \frac{ v R'}{ N}\, ,
\end{equation}
again yielding $v/f \ll 1$ assuming that the KK scale is of order $1/R' \sim 1$ TeV. 
Thus the basic RS radion can not be successfully used to replace the higgs. On the other hand, due to the very large kinetic term for the radion its mass will be even further suppressed compared to the KK mass scale:
  \begin{equation}
m_{dil}^2= \frac{16}{N R'^2} \left( v_1 \sqrt{-\delta a} - \frac{\delta a}{2} \right) \epsilon + O(\epsilon^2) \, .
\end{equation}
Once $\delta a$ is tuned to obtain the right minimum $\sqrt{-\delta a} \sim {\cal O}(v_1)$ we can see that we get a radion mass that is significantly lighter than the KK mass scale in the theory 
as explained in~\cite{RattazziZaffaroni,CGK,CGRT,GW2}:
\begin{equation}
m_{dil} \sim  M_{KK} \frac{2 v_1 \sqrt{\epsilon}}{\sqrt{12 (M_* R)^3}}\, .
\end{equation}


\section{Conclusions}
The idea of linking EWS breaking and spontaneous conformal breaking has a long and varied history.  There is undeniably a theoretical appeal to realizing the physical higgs boson of the SM as a pseudo-Goldstone boson of spontaneous conformal breaking.  What we have shown in this paper is that, depending on assumptions about how the SM sector is embedded in the conformal sector (especially with regards to compositeness or partial compositeness), a dilaton can in principle have the couplings observed at the LHC for the 125 GeV higgs-like resonance and be consistent with EWPT. However these appealing prospects come with a stiff price: fine-tuning and strong dynamical assumptions.  In order to get a light dilaton it is not enough to have a conformal theory and simply introduce some small breaking parameter.  One must ensure that the breaking scale, $f$,  is stabilized far enough below the UV scale so that there is a large range of energies with approximate conformal behavior.  The quartic dilaton coupling must be reduced by some kind of tuning from its NDA value of $16\pi^2$ to much less than one.  The operator that breaks the conformal symmetry must be arranged to be almost marginal and its coupling must remain perturbative over the conformal range and well bellow $f$.  In the simplest non-supersymmetric model where all of these conditions can be arranged for, the GW stabilized RS model, one can produce a light dilaton, however despite all the
care that has been taken there still remains an overly large kinetic term for the radion/dilaton which suppresses its couplings to the SM sector and thus the dilaton fails to imitate the SM higgs.
It remains an open question whether there exists a consistent, non-supersymmetric model that has a light enough dilaton that can also mimic the SM higgs boson couplings.
A further daunting requirement would be to arrange all this without fine-tuning. 

\section*{Acknowledgements}

We thank Matt Reece for stimulating conversations about the dilaton mass. We also thank Kaustubh Agashe, Zohar Komargodski and Alex Pomarol  for useful discussions.  We thank Zackaria Chacko for informing us of his upcoming paper on this subject.  We thank the Aspen Center for Physics for its hospitality while this work was in progress. 
B.B. is supported in part by the ERC Advanced Grant no.267985, ``Electroweak Symmetry Breaking, Flavour and Dark Matter: One Solution for Three Mysteries'' (DaMeSyFla), and by the MIUR-FIRB grant RBFR12H1MW.
 C.C. and J.S. are supported in part by the NSF grant PHY-0757868.  J.H. thanks Cornell University for hospitality during the course of this work.  J.H. is supported in part by DOE grant number DE-FG02-85ER40237.  J.T. was supported by the
Department of Energy under grant DE-FG02-91ER406746.


\end{document}